\documentstyle[psfig,preprint,aps,prl]{revtex}

\begin{document}
\draft

\title{Quantifying Nonstationary Radioactivity Concentration
Fluctuations Near Chernobyl: A Complete Statistical
Description}

\author{G. M. Viswanathan$^{1,2,3}$,
S, V. Buldyrev$^{1}$, E. K. Garger$^{4}$, V. A. Kashpur$^{4}$,\\
L. S. Lucena$^{2}$, A. Shlyakhter$^{5}$, H. E. Stanley$^1$ and
J.Tschiersch$^6$}

\address{$^{1}$Center for Polymer Studies and Department of Physics,
Boston University, Boston, MA 02215}

\address{$^{2}$International Center for Complex Systems and 
Departamento de F\'\i sica
Te\'orica e Experimental, Universidade Federal do Rio Grande do Norte,
59072-970, Natal--RN, Brazil}

\address{$^{3}$Departamento de F\'\i sica, Universidade Federal de Alagoas,
57072-970, Macei\'o--AL, Brazil}

\address{$^{4}$Institute of Radioecology, Ukrainian Academy of
Agricultural Sciences, \\
252033 Kiev, Tolstoy St.~14, Ukraine}

\address{$^5$Lyman Laboratory of Physics, Harvard University,
Cambridge, MA 02138}

\address{$^6$Inst. of Radiation Protection, GSF--Nat'l Center 
for Environment and Health, D--85764, Neuherberg, Germany}
\date{\today}

\maketitle\vspace{-1cm}

\begin{abstract}

We analyze nonstationary $^{137}$Cs atmospheric activity concentration
fluctuations measured near Chernobyl after the 1986 disaster and find
three new results: (i)~the histogram of fluctuations is well described
by a log-normal distribution, (ii)~there is a pronounced spectral
component with period $T=1$~y, and (iii)~the fluctuations are long-range
correlated.  These findings allow us to quantify two fundamental
statistical properties of the data---the probability distribution and
the correlation properties of the time series.  We interpret our
findings as evidence that the atmospheric radionuclide resuspension
processes are tightly coupled to the surrounding ecosystems and to large
time scale weather patterns.

\end{abstract}

\section{Introduction}

Chernobyl's No.~4 reactor was completely destroyed on 26 April 1986 by
explosions that blew the roof off the reactor building and released
large amounts of radioactive material into the environment, particularly
during the first ten days. The discharge included over 100
radionuclides, mostly short-lived, but the radioactive isotopes of
iodine and cesium were of special radiological relevance from a human
health and environmental standpoint.  The cesiums, with lifetimes of the
order of tens of years, have long-term radiological impact.  Especially
important is $^{137}$Cs, a $\beta$- and $\gamma$-emitter with a
half-life of 30.0 years (specific activity 87 Ci/g). Its decay energy is
1.176 MeV; usually divided by 514~KeV $\beta$ and 662~KeV $\gamma.$ It
comprises some 3--3.5\% of total fission products. (It is the primary
long-term $\gamma$ emitter hazard from fallout, and can potentially
remain a hazard even for centuries.)  Radioactive material from the
reactor plant was detectable at very low levels over practically the
entire Northern Hemisphere.  Motivated by the need for reliable
inhalation dose rate estimates of radioactive isotopes, daily air filter
samples of aerosol were collected in 1987---1991 using high-volume
samplers \cite{original}. Filters exposed to aerosols were
pressed into discs and analyzed by $\gamma$-spectrometry to find the
activity concentrations of atmospheric $^{137}$Cs.  Previous analyses by
Garger {\it et al.} \cite{garger} revealed an exponential decrease of
the absolute values of the activity concentration with time. The
decrease was faster than expected assuming radioactive decay only,
suggesting that mechanisms such as vertical migration in soil, run-off
with rain, or melting water and snow cover may be responsible for the
reduction of the activity concentration.  Indeed, several experts had
apparently pointed out at the I.A.E.A. Experts' Meeting in Vienna,
August 1986, that the decay rate is faster than should be observed if
radioactivity were the only process involved \cite{vienna86}. This fast
decay rate was a criticism of the prediction of future doses by
Pavlovskij which he has since apparently accepted \cite{pavlovskij}.

It is crucial, when analyzing such data, to know how the data were
collected. (For example, the activity in a town near the power plant
will be of a different nature from data collected in a nearby forest,
etc.)  The data we analyze were taken starting in June 1987 with
high-volume samples in the town of Pripyat, located 4~km from the
nuclear power plant in the northwest direction. The resuspension of
particulates deposited on the land surface following atmospheric
releases of radioactive material from Unit 4 of the Chernobyl Nuclear
Power Station in 1986 has been studied for the past six years (see,
e.g., ref. \cite{original}). In that region, resuspension is a result of
the action of wind blowing across the terrain and from mechanical
disturbances such as the movement of motor vehicles and the activity of
farm equipment. In addition, the data gathered near Pripyat reflect
major disturbances such as the demolition of buildings and the burial of
the highly radioactive ``red'' forest. These activities took place prior
to 1989. Enhanced resuspension may also have resulted from
major building and forest fires within the 30~km exclusive zone and
decontamination works.

The fluctuations in the activity concentrations are highly nonstationary
in both space and time. It is well known, for example, that in the
summer the radioactivity is almost threefold higher when passing near
woodlands than when in open country. This increase is attributed to
trees picking up radioactivity, and will result in a seasonal
fluctuation with a period of $T=1$~y.

Previous attempts to quantify the experimentally measured probability
distributions of the concentration fluctuations were limited to
statistical characteristics important for radioecological and
radiological estimations and were inconclusive because the data did
not fit the log-normal, exponential, or gamma distributions. Hence,
statistical approaches that apply fractal concepts to the data
analysis have been useful. Hatano and Hatano, for example, proposed a
model with fractal fluctuation of wind speed that successfully
reproduces data of the aerosol concentration measured near Chernobyl
over a decade \cite{hatanohatano97}. They also are able to reproduce
the time dependence of the resuspension factor and obtain values of
the fitting parameters that provide important information on the
emission quantity and removal processes of nuclides from the
accident. Using their model, they predicted a power law decay of the
average aerosol concentration with an exponential cutoff: $\langle
C(t)\rangle\sim e^{-t/\tau}t^{-4/3}$, and temporal correlations that
decay as a power law $\sim t^{-2/3}$.  Here we apply a different
statistical description of the data and show that after removing the
exponential trend in the data, the residual fluctuations are well
described by a log-normal distribution.  Specifically, we find that
the logarithm of the concentration is a linearly decaying function of
time with fluctuations that are characteristic of long-range power law
correlated Gaussian noise, with temporal correlations decaying as
$t^{-0.3}$, i.e., much slower than the model \cite{hatanohatano97}
predicts.  On the other hand, we were able to find no indication of
the power law behavior of the average concentration predicted in
\cite{hatanohatano97}.

\section{Methods and Results}

\begin{figure}
\vspace{.2in}

\bigskip
\centerline{\psfig{figure=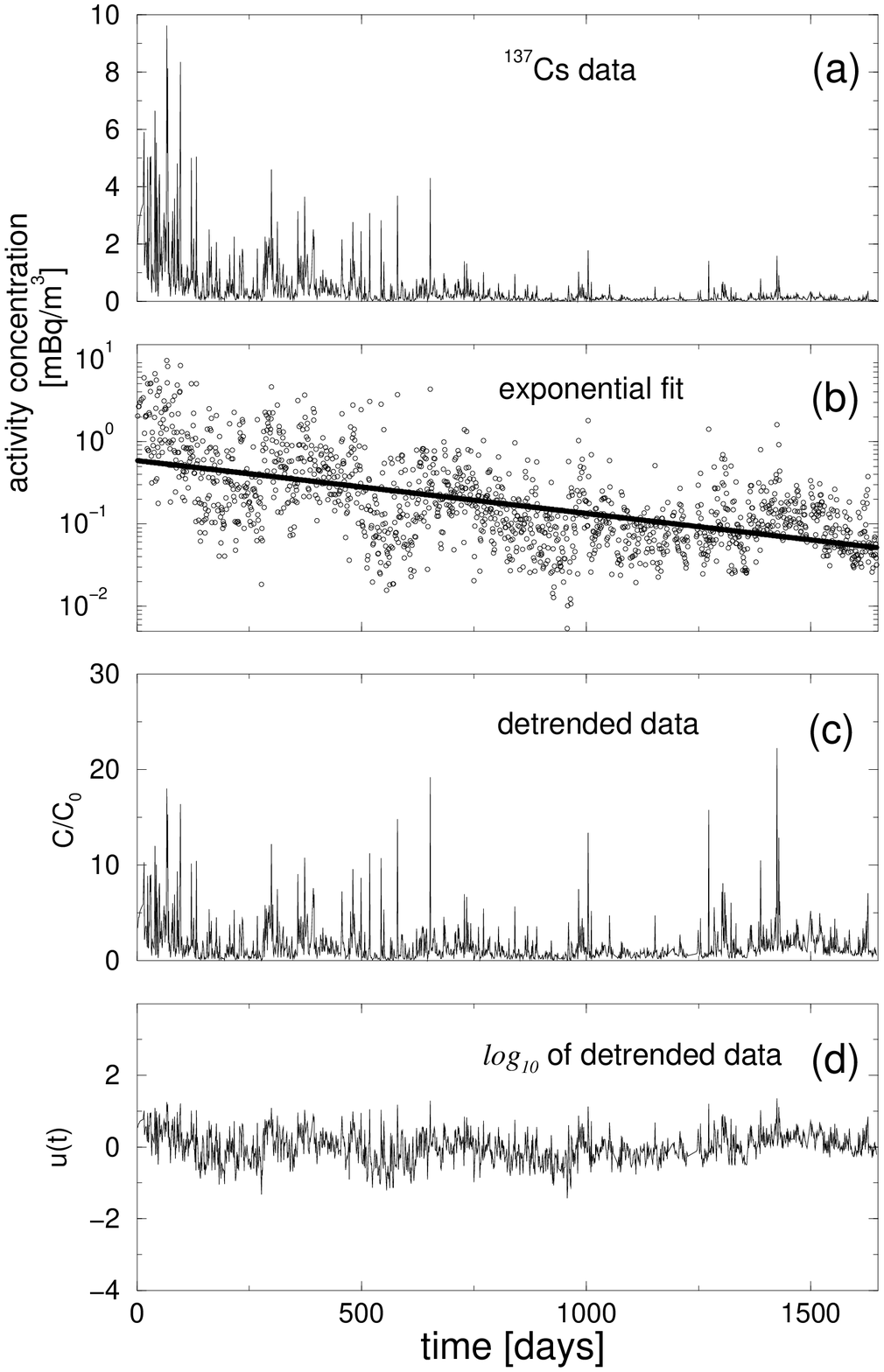,height=6truein}}
\caption{ (a) $^{137}$Cs activity concentration $C(t)$ measured after
the Chernobyl disaster as a function of time $t$ in days.  (b)~Semi-log
plot of the data and the exponential best fit $C_0(t)\sim \exp[-t/677],$
obtained by regression (solid line).  (c)~Fluctuations after correcting
for the exponential decay.  The data $C(t)$ have been divided by the
exponential fit $C_0(t)$.
(d)~Logarithms $u(t)\equiv \log_{10}[C/C_0]$ of these data.  The
nonstationarity apparent to the naked eye in (a)~appears eliminated in
(c)~and~(d).}
\label{intro}
\end{figure}

The original activity concentration data $C(t)$ as a function of time
$t$ are shown in Fig.~\ref{intro}(a). We apply regression to the data to
obtain the exponential best fit (Fig.~\ref{intro}(b))
\begin{equation}
C_0(t)=A\exp(-t/\tau),
\end{equation}
and empirically find $A=0.588$~mBq/m$^3$ and $\tau=677$~days.  We
``detrend'' the data $C(t)$ by dividing it by the exponential fit
$C_0(t)$ (Fig.~\ref{intro}(c)). We thus obtain a dimensionless time
series that represents relative concentration fluctuations away from
the exponentially decaying geometric mean value.  Fig.~\ref{intro}(d)
shows the logarithms $u(t)\equiv \log_{10} [C(t)/C_0(t)]$ of these
fluctuations.  (We choose a logarithmic measuring unit for the same
reason that acidity and sound intensity are measured in pH in dB
respectively.)  The nonstationarity (of the mean) in the data,
apparent to the naked eye in Fig.~\ref{intro}(a), appears eliminated
in Fig.~\ref{intro}(c) and Fig.~\ref{intro}(d).

\begin{figure}
\vspace{.2in}

\bigskip
\centerline{\psfig{figure=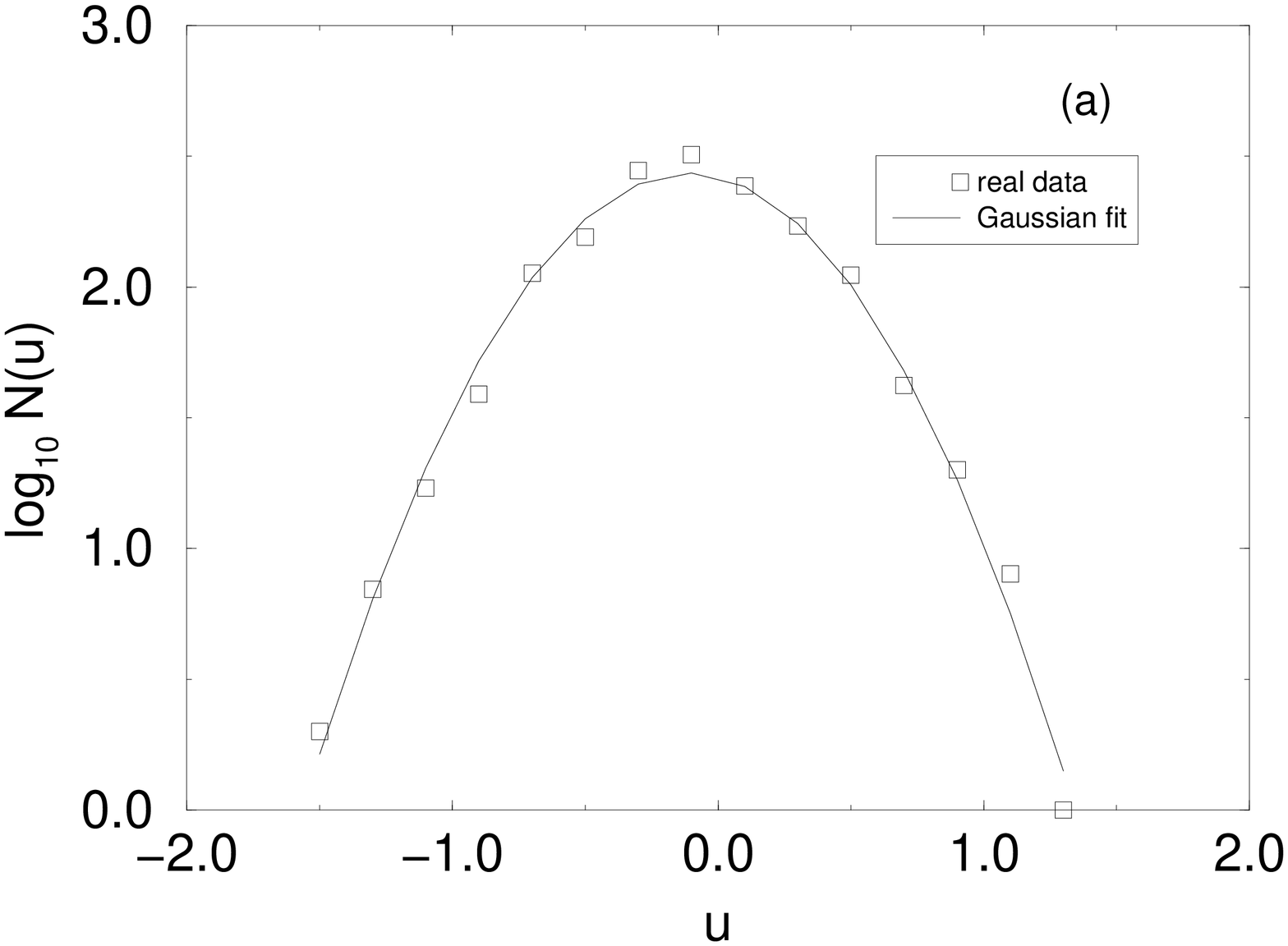,angle=-90,width=3.5truein}}
\centerline{\psfig{figure=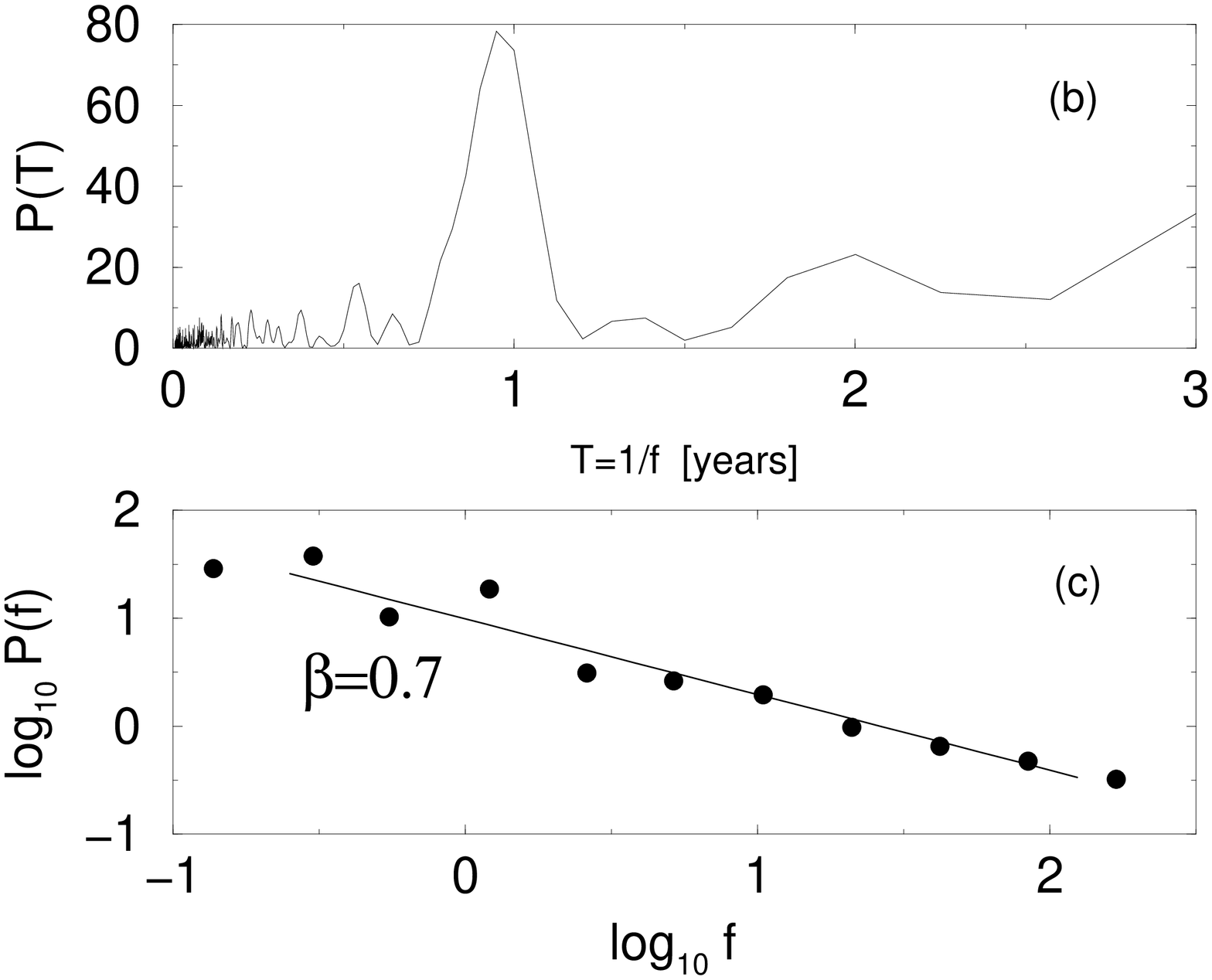,angle=-90,width=3.4truein}}
\smallskip
\centerline{\psfig{figure=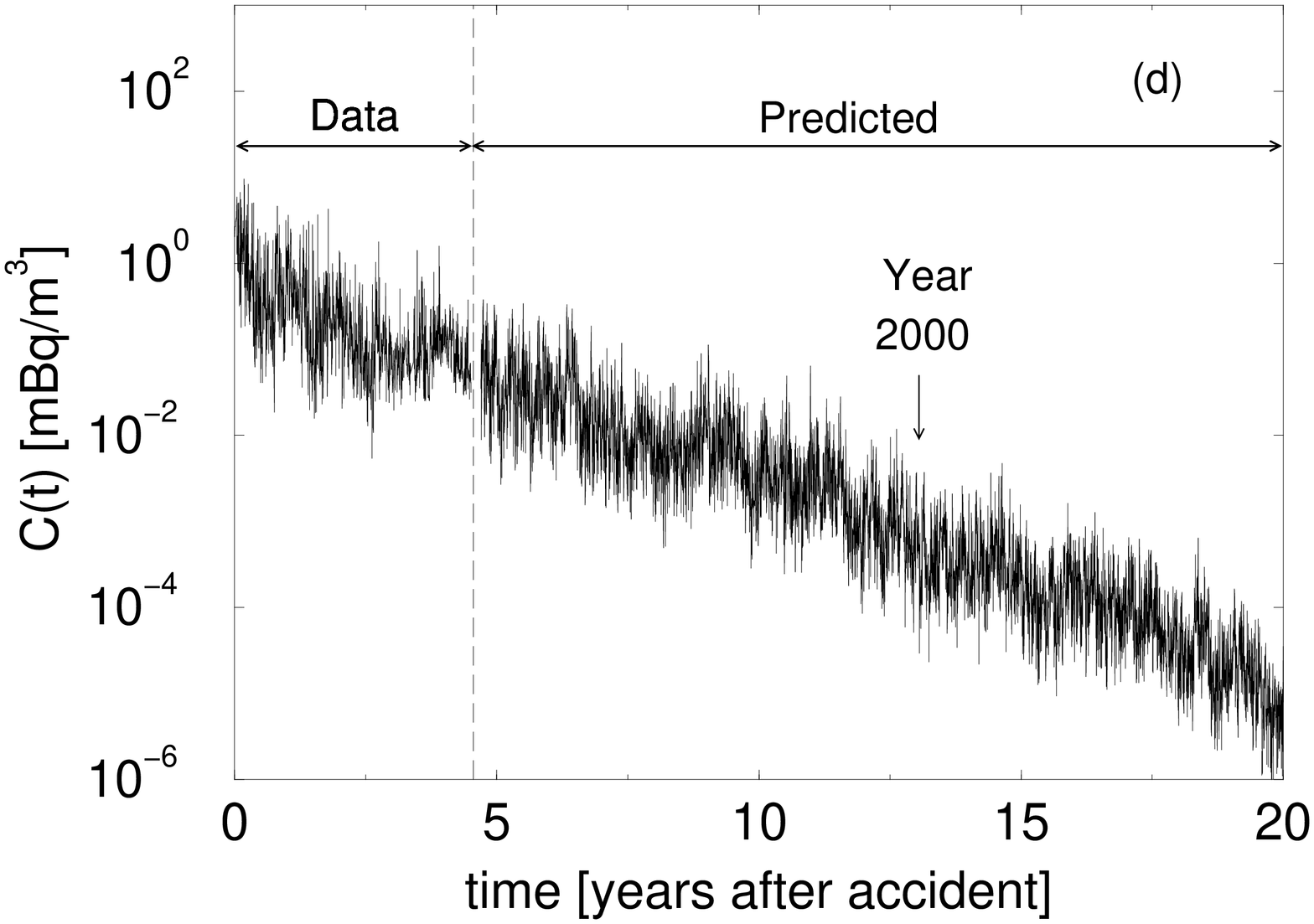,angle=-90,width=3.5truein}}

\caption{ (a)~Semi-log plot of the histogram of $u(t)$ along with a
Gaussian fit, showing that $u$ closely follows a Gaussian distribution
(a parabola in semi-log plot).  The fluctuations in $C(t)$ are thus seen
to follow a log-normal distribution, after being corrected for the
observed exponential decay.  
(b)~Normalized power spectrum $P(f)$ of $u(t)$ as a function of the
period $1/f$ in years. Due to the lack of data for up to 7\% of all
days, the spectrum shown has been computed using a Fourier transform
algorithm that is immune to the ``missing data'' problem as discussed in
the text. The pronounced 1~y and 2~y spectral components suggest that the
fluctuations are influenced by environmental factors.  (c)~Double log
plot of $P(f)$ as a function of frequency measured in years$^{-1}$. We
have applied logarithmic binning to smooth the spectrum by averaging
over windows which double in size.
We find that $\beta$~($\approx 0.7$) is significantly different from
zero, so we conclude that the activity data $C(t)$ are long-range
correlated and log-normally distributed around the exponentially
decaying function $C_0(t)$.  (d)~Semi-log plot of activity
concentrations as a function of time after the nuclear accident, up to
20~years.  Experimental data are shown for all the dates
when they are available. For future dates when data are not available,
we use the model discussed in the text to predict the long-term
behavior.  The predicted radiation levels for the future (12 years after
the accident onwards) is below the generally accepted safe level.}

\label{spectrum}
\label{model}
\label{histo}
\end{figure}

We apply histogram techniques to find the probability distribution of
$u(t).$ Fig.~\ref{histo}(a) shows that $u(t)$ (Fig.~\ref{intro}(d)) is
Gaussian distributed. Therefore, the signal in Fig.~\ref{intro}(c) is
log-normally distributed.  Since log-normal distributions often arise
from underlying multiplicative processes, we study correlations in
$u(t).$

The power spectrum of $u(t)$ cannot be computed using the fast Fourier
transform (FFT) algorithms because the data are missing on up to 7\%
of all days.  We calculate the power spectrum $P(f)$ of the sequence
$u(t)$ using a Fourier transform algorithm that can correctly treat
the ``missing'' data.  The ``Lomb normalized power spectrum''
\cite{lomb}, $P(f)$, often used when the data are unevenly sampled or
missing, is shown in Figs.~\ref{histo}(b) and~(c) on a double log
scale.  (The Lomb spectrum gives an estimate of the harmonic content
of a data set for frequency $f$ that is based on a linear
least-squares fitting to the function $A\sin(2\pi ft)+B\cos(2\pi ft)$
\cite{lomb,numerical}.)  We find prominent 1~y and 2~y spectral
components in the data. This finding suggests that the fluctuations
are due in part to environmental factors.  Since the environment has
``memory'' (long-range correlated behavior), we next test for
long-range correlations in the data.  The linear behavior of the
log-log plot of
\begin{equation}
P(f)\sim f^{-\beta}\;\;,
\end{equation}
indicates the presence of scale invariant behavior. The scaling exponent
$\beta\approx 0.7$ indicates that the data are long-range correlated,
since the autocorrelation function decays in time with exponent
$\gamma=1-\beta$ for $0 < \beta < 1.$ Note that $\beta=0$ for
uncorrelated white noise and $\beta=1$ corresponds to $1/f$-type noise.
We have verified our finding of long-range correlations using other
methods and find consistent results.

\section{Model}

We model the long-term behavior of the activity concentration
fluctuations. We first generate a sequence $w(t)$ of long-range
correlated Gaussian noise with $\beta=0.7$ (the value of $\beta$
observed experimentally).  We then exponentiate this sequence to
obtain a second sequence $x(t)$ that is log-normally distributed as
well as long-range correlated. The normalization of $x(t)$ is such
that the mean and variance are fixed to be identical to those of the
sequence $u(t)$ obtained from real data.  Next, we multiply the
sequence $x(t)$ by the exponential best fit function $C_0(t)$ found
empirically, thereby obtaining the model sequence $C_M(t)$
(Fig.~\ref{model}(d)). The model sequence $C_M(t)$ and the original
data $C(t)$ are characterized by the same exponential trend,
log-normal distribution, and power spectrum.  The predicted activity
concentrations never rise above 10$^{-2}$~mBq/m$^{-3}$ after 1998,
even for extremely large fluctuations. The prediction is encouraging
from a human health and environmental perspective, since this level is
considered to be safe.

\section{Discussion}

The log-normal distribution \cite{bencala} could be related
\cite{garger} to local resuspension of the radionuclides by wind and
transport from distant areas. Additional support for the possibility
that local resuspension is the leading factor determining the observed
fluctuation arises (i) from the presence of the pronounced 1~y spectral
component and (ii) from the observed long-range correlations, as we next
discuss:

\begin{itemize}

\item[{(i)}] A pronounced 1-year periodicity can be attributed to
large-scale seasonal environmental changes, such as snow melting, as
well as to human agricultural activity, such as plowing. Both yearly
processes can release radioactive elements accumulated in soil.

\item[{(ii)}] Long-range correlations with shorter characteristic
times can be generated by large-scale atmospheric turbulence. It is
possible that resuspension of radionuclides due to changing weather
conditions \cite{bunde} give rise to long-range correlated
fluctuations around the exponentially decaying average activity
concentrations, $C_0(t)$.  However, it is clear, that it is not only
weather that is responsible for the observed long-range correlations
in the activity levels. Ecosystems consist of a large number of
reservoirs (e.g., forests, lakes, swamps and agricultural fields) of
various sizes that can repeatedly absorb and release radionuclides
into the environment \cite{eco}. It is well known since the works of
Hurst \cite{hurst} that the level of water in lakes is characterized
by power law decaying fluctuations with the Hurst exponent $H\approx
0.7$ \cite{hurst}, The amount of radionuclides released in a unit of
time should be proportional to the derivative of these fluctuations
and hence should have a power spectrum scaling exponent $\beta=2H-1$
\cite{vicsek}, which is in qualitative agreement with our
measurements.  The value of $\beta\approx 0.7$ found by us is larger
than the value $0.4$ obtained from Hurst measurements and the
discrepancy may be partially related to the fact that the high
frequency part of the power spectrum (see Fig.~\ref{spectrum}(c)) has
smaller slope $\beta\approx 0.5$. Similar values of $\beta\approx 0.5$
as well as log-normal distributions of fluctuations were obtained for
river discharges \cite{Pelletier97}.  Moreover, the radioactive
release is related not only to lakes but also to other parts of the
ecosystem, which may have different Hurst exponents.  Indeed, the
influence of weather (and possibly forest fires) on monthly mean
$^{137}$Cs concentrations provides an illustration of how the
ecosystem can absorb and release radionuclides \cite{garger-eco}.  The
power law behavior of the power spectrum indicates that the temporal
correlations decay as a power law $\sim t^ {-\gamma}$, with
$\gamma=1-\beta$ \cite{Peng93}. Hence our analysis yields
$\gamma=0.3$.  The model of wind resuspension used in
\cite{hatanohatano97} also predicted power law decaying correlations
but with $\gamma=2/3$, indicating a much faster decay than we observe.
Indeed, such slowing down of the correlation decay with respect to the
pure atmospheric activity is often observed in the the response of
ecosystems such as forests and ground waters \cite
{Pelletier97,Lange99}.

\end{itemize}

\section{Conclusion}

In summary, we have attempted to find a complete statistical
description of the data.  We have succeeded in quantifying both the
probability distribution as well as the correlation properties of
$C(t)$---the two most fundamental statistical properties of time
series.  As with refs. \cite{hatanohatano97,hatanohatano2}, the values
obtained for the fitting parameters provide important information on
the emission quantity and removal processes of radionuclides from the
accident site.  The differences between our results and those of
\cite{hatanohatano97,hatanohatano2} may indicated the significant role
of the ecosystems in the radionuclides resuspension. One important
conclusion is that the observed long-range correlations are not solely
due to the atmospheric turbulence but are strengthened by the
accumulation of radionuclides by ecosystems, such as forests and
ground waters.

\subsubsection*{Acknowledgments}

We thank C. Argolo, M. Barbosa, T.~Keitt, S.~Lovejoy, M.~L.~Lyra,
R.~Mantegna, R.~Nowak and T.~J.~P.~Penna for discussions. We thank the
Brazilian agency CNPq and NSF for support.

\end{document}